\documentclass{aa_nohead}

\usepackage{epsf}

\input epsf

\begin{document}

\thesaurus{(08.14.2; 08.23.1; 02.03.3; 02.08.1; 02.14.1)}
\title{Two-Dimensional Simulations of the Thermonuclear Runaway in an Accreted Atmosphere of
a C+O White Dwarf}
\author{A. Kercek \inst{1} \and W. Hillebrandt \inst{1} \and J. W. Truran \inst{2}}
\institute{Max Planck Institut f\"ur Astrophysik,
Karl-Schwarzschild-Strasse 1, D-85740 Garching, Germany
\and Department of Astronomy and Astrophysics, Enrico Fermi Institute,
University of Chicago, Chicago, IL 60637, USA}
\date{Received <date>; accepted <date>}
\maketitle
\markboth{A. Kercek et al.: Two Dimensional Simulation of a TNR on a
CO White Dwarf}
{A. Kercek et al.: Two Dimensional Simulation of a TNR on a CO White Dwarf}
\begin{abstract}
We present the results of two-dimensional calculations of turbulent 
nuclear burning of hydrogen-rich material
accreted onto a white dwarf of 1.0 M$_{\odot}$. The main aim of the
present paper is to
investigate the question as to whether and 
how the general properties of the burning are
affected by numerical resolution effects. In particular, we want to see
whether or not convective overshooting into the surface layers of the C+O
white dwarf can lead to self-enrichment of the initially solar 
composition of the hydrogen-rich envelope with carbon and oxygen from the 
underlying white dwarf core. 

Our explicit hydrodynamic code is based on the
PPM-method and computes the onset of the thermonuclear runaway 
on a Cartesian grid. Only part of the white dwarf's surface is covered
by the computational grid and curvature effects are ignored.
In contrast to previous works we do not observe fast mixing of carbon and
oxygen from the white dwarf's surface into the envelope by violent overshooting 
of large eddies. 
The main features of the flow fields in our simulations are the appearance
of small persistent coherent structures of very high vorticity 
(and velocity)  compared to the
background flow. Their typical  linear scales are about 10 to 20 grid zones
and thus their physical size depends on the numerical resolution, i.e,
their size decreases with increasing resolution. For the early phase
of the thermonuclear runaway (TNR)
they dominate the flow patterns and result in very little
overshoot and mixing. Only at late times, after steady slow mixing and
with increasing nuclear energy production, do these structures become
weak, but show up again once hydrogen has mainly been burnt and the
energy generation rate drops. 

On the other hand, there are no big
differences between high and low resolution simulations, as far as the
overall properties of the TNR are concerned.
The two simulations which are presented here 
show only moderate differences in spatially
integrated quantities such as laterally averaged temperature, energy
generation rate, and chemical composition. We have not expanded both
simulations equally long, but  
for the physical time under consideration the major difference seems
to be that the highly resolved simulation is a bit less violent. 
In conclusion, we do find some self-enrichment, but on time-scales much
longer than in previous calculations. 


\end{abstract}

\section{Introduction}

It is generally accepted that the outburst of a classical nova
is caused by a thermonuclear runaway (TNR) 
in the accreted hydrogen-rich envelope on top of a white dwarf in a close binary
system (\cite{Starr89} 1989; \cite{Starr93} 1983;
\cite{Starr95} 1995; \cite{Truran82} 1982; \cite{Truran90} 1990).
Detailed one-dimensional spherically symmetric models of the runaway,
based on realistic nuclear reaction rate networks but using the mixing-length
theory (MLT) of convection, have been studied extensively 
(\cite{Starr74} 1974,1985; \cite{Prialnik78} 1978;
\cite{MacD80} 1980), and the results
are in good agreement with observational data, such as the
total amount of energy released and the metallicity 
and the abundances of the expelled envelope material.
 
However, using the MLT to model the runaway
is very questionable, as can be seen from simple arguments. 
In general terms, the physics of the TNR can be described as follows.
For a white dwarf accreting matter from a companion star of roughly solar 
composition, during the accretion process the
temperature at the base of the envelope and the nuclear 
energy release increase steadily but
slowly. Once the temperature exceeds about
$2 \cdot 10^7$ K heat conduction can no longer transport the energy
released by nuclear reactions fast enough and the envelope becomes
unstable to convection. However, because of electron degeneracy
the nuclear energy released does not go into lifting off the envelope
at this stage, but the  
temperature increases further and the TNR proceeds until
the electron degeneracy is removed and an overall expansion sets in. 
At the onset of the violent burning phase 
the temperature at the bottom of the envelope is about
10$^8$ K and it lasts for several minutes. Energy production is dominated by the
$\beta$-limited CNO-cycle since the temperatures
never exceed 4$\cdot$10$^8$ K (see, e.g., \cite{Wallace81} (1981)). Therefore, 
the energy production per unit mass is  independent of temperature and
density and is just a linear function of metallicity (\cite{Wallace81}
1981).
The limitation of the nuclear energy generation rate is the reason
why hydrogen burning on top of a white dwarf is different from, say,
the carbon-burning deflagration wave inside a Chandrasekhar-mass white
dwarf. In the latter case, due to the temperature sensitivity of the 
energy generation rate and the small thermal width of the burning
region a flame forms and is passively advected by the turbulent
motions. Combustion is said to be in the ``flamelet'' regime.
In contrast, in the phase of
CNO-limited hydrogen burning on top of a white dwarf
burning is always slow compared to convective turn-over and combustion
takes place in the so-called ``well-stirred'' regime for which either
direct numerical simulations have to be carried out or pdf-methods
have to be applied. In either case, the treatment of convection is
crucial and requires special care because convection dominates over
burning.

A second aspect which makes reliable numerical simulations of novae
very difficult is the fact that    
due to diffusion and coupling to hydrodynamic flows 
(especially convective overshooting 
(\cite{Woosley86} 1986)) there can be a significant increase of
the energy production rate by mixing C and O from the white dwarf core
into the envelope. The $\beta$-limited CNO-cycle operates no longer in
a strict equilibrium, but due to the increase of the C and O
abundances in mixed zones
the energy generation rate is time-dependent and couples to the mixing
processes in a very non-linear manner: Violent motions can cause
mixing which in turn may make the motions even more violent.  
It was this effect that lead to fast self-enrichment and a rapid
temperature rise in the two-dimensional direct simulations 
of Glasner, Livne \& Truran (1997).

A third question concerns the ignition process.
One dimensional models with spherical symmetry consider a global and simultaneous
ignition at the bottom layer of the envelope (global TNR or GTNR). 
However, according to Shara (1981,1982) also
a local TNR (LTNR) is possible, because of temperature inhomogeneities arising from magnetic
fields, rotation and non-spherically symmetric accretion. Since  
it would take years to spread the combustion front
from a LTNR over the white dwarf's surface diffusively 
by radiative transport and thermal conduction (\cite{Shara81_82} 1981,1982),
Fryxell \& Woosley (1982) discuss that heat transport by mass motions as a multidimensional
effect dominates over diffusion. For typical situations at peak temperature
they find laminar burning velocities of about 
$v_{\rm lam}$ = 30 cm/s only. In contrast, estimating the effective
burning velocity caused by small-scale
turbulence motions to be
\begin{equation}
v_{\rm burn} = v_{\rm lam} + \left( \frac{h_{\rm p}v_{\rm c}}{\tau_{\rm burn}} \right),
\end{equation}
where $h_{\rm p}$ is the pressure scale hight, and $v_{\rm c}$ the typical convective velocity,
they obtain $v_{\rm burn}$ $\sim$ 10 km/s for a typical nova model.
So again, multi-dimensional effects can be crucial for the TNR.

To date, not very many multi-dimensional studies exist which deal with
the various questions raised above. Shankar, Arnett \& Fryxell (1992);
Shankar \& Arnett (1994); Glasner \& Livne (1995);
Glasner, Livne, \& Truran (1995) and Glasner, Livne, \& Truran (1997) 
were the first to carry out two-dimensional direct simulations of TNRs.
In the most recent computation (\cite{Glasner97} 1997) no initial
enrichment of carbon and oxygen was necessary to initiate  a rapid TNR
which, likely, will lead to an outburst similar to that of a
classical nova. In their case violent large eddies were responsible
for considerable mixing of envelope and core matter. Their
computations were performed with the (implicit) VULCAN-code 
(\cite{Livne93} 1993) in 
spherical coordinates for a slice of 0.1 $\pi$ radians with reflecting
boundaries at the bottom and on both sides.  

Because their results are so important and because sometimes numerical
effects cannot be separated from physical ones easily
we decided to perform independent simulations based upon a different
hydro-code (PROMETHEUS; \cite{Fryxell89} (1989)).
Moreover, we decided to ignore curvature
effects and to represent the surface layers of the white dwarf as well
as its envelope by a plane-parallel sheet. The advantage of this
approach is that we can use periodic boundary conditions and thereby
avoid common problems of numerical simulations of free convection,
namely that reflecting boundaries may act like a containment and may affect
the flow patterns in an unphysical way. In spherical coordinates
with the polar axis singled out one cannot use periodic boundary
conditions. We will come back to this point later in Sec. 4 when we
interpret our results and compare them with \cite{Glasner97} (1997).

In the following we present the first results of our 
two dimensional (2-d) simulations starting from the same initial
conditions as \cite{Glasner97} (1997). We performed the simulations with two different Cartesian
grids covering the same domain as in Glasner, Livne \& Truran (1997). The first grid
has approximately the same numerical resolution as was used by 
Glasner, Livne \& Truran (1997),
 and the second one uses 5 times more zones in each direction.
In Sec. 2 we give a description of our numerical method, including the hydro-code,
the nuclear reaction network, and details of the initial conditions.
In Sec. 3 we present the results, and Sec. 4 is devoted to a discussion
and, in particular, to a comparison with the
work of \cite{Glasner97} (1997).
The fact that according to our work 2-d simulations predict that 
during most of the TNR large scale motions do not dominate the
evolution gives us hope that going to the full 3-d problem may not 
change all our conclusions. 3-d calculations which will proof or disproof
this conjecture are under way and will be published in a subsequent
paper.
 
\section{The code and initial conditions}

%$^{14}O(e^+,\nu )$$^{14}N(p,\gamma )$$^{15}O(e^+,\nu )$$^{15}N(p,\alpha )$$^{12}C(p,\gamma )$$^{13}N(p
%,\gamma )$$^{14}O$.

Our calculations were performed with a slightly modified version of 
PROME\-THEUS, an Eulerian code based on a second order
Godunov method with piecewise parabolic interpolation of the conserved quantities
for solving the hydrodynamic equations, as worked out by
Colella \& Woodward (1984) 
and \cite{Fryxell89} (1989). Since this code is used frequently by various
groups (see, e.g., M\"uller (1994,1997)
\cite{Shankar94} (1994) and Niemeyer \& Hillebrandt (1995)) 
we do not give a detailed
description here or provide any numerical tests. We want to mention in
passing, however, that we had to change a few things in the original 
version of the code to make it applicable to the problem under consideration. First, a
challenging problem was to keep the initial model stable over several
thousand dynamical time scales if nuclear reactions were suppressed.
The second problem was the presence of huge gradients at the core-envelope
interface which lead to unacceptable numerical diffusion.

In order to get a stable initial model we solved the equations of 
hydrostatic equilibrium for the pressure with a Runge-Kutta-solver, using 
the density and 
composition profiles of the initial model of \cite{Glasner97} (1997), which
will be described in more detail later in this chapter, and our equation of state.
Next we mapped this
density distribution on to the grid required
by PROMETHEUS. The deviations from the original
model of Livne and Glasner were less than 1 \% in pressure and density. 

The huge gradients at the core-envelope interface required special attention
since gravity in most versions of PROMETHEUS is included by
calculating the gravitational acceleration {\it after} each hydro time step.
Since PROMETHEUS is a conservative second order Godunov scheme one cannot
impose hydrostatic equilibrium directly. Consequently, pressure gradients
are not balanced exactly by gravity. The accuracy one can reach is
about 0.1 \% which causes a net acceleration for the core-envelope-interface 
resulting in a systematic slow but significant outward motion of the interface.

To overcome this problem we have treated the time evolution
of the gravitational acceleration in a more symmetric way.
It was included by calculating its action prior to and after the hydro
timesteps with weighting factors f and (1 - f), respectively. We found
that for f = 0.12 the interface stayed in place for several thousand
dynamical timescales, after the model had been relaxed, with nuclear
reactions being turned off. 

In fact, we found that these modifications of the original PROMETHEUS scheme 
were significant but not crucial. As a test, we performed a 2-d
simulation with nuclear reactions 
turned on for f=0 where the interface was moving 
outwards with the same speed as in the 1-d case without nuclear burning.
We realized that except for the 
moving of the transition region and some unphysical
strong heating at the inner reflecting boundary the flow fields did
not change by much. The main difference was a slightly higher energy 
production rate for the f=0 case since numerical diffusion 
mixed additional core material into the
burning region of the envelope. 
 
Finally, the code was changed to run efficiently on massively parallel
computers such as the CRAY T3E by means of explicit message passing.

In our version of PROMETHEUS 
nuclear reactions are incorporated by solving 
together with the hydrodynamics a nuclear reaction network including 
12 nuclear species, i.e., $^1$H, $^4$He, $^{12}$C, $^{13}$C, $^{13}$N, $^{14}$N,
$^{15}$N, $^{14}$O, $^{15}$O, $^{16}$O, $^{17}$O, and $^{17}$F, linked
by reactions described in \cite{Wallace81} (1981). The reaction rates 
are taken from \cite{Thielemann96} (1996). 
Following \cite{Mueller86} (1986), we solve the network 
equations and the 
energy source equation simultaneously to avoid numerical instabilities. 

For a given density $\rho$ the network equations and
the energy source equation have the following form: 
\begin{eqnarray}
\dot{Y}_i - F(Y_j,T) & = & 0 \\
\dot{\epsilon} - H(Y_j,T) & = & 0,
\end{eqnarray}
where $Y_i$ are the nuclear abundances,  $\dot{\epsilon}$ is the energy
production rate per unit mass, i,j = 1...N, and N is the number of nuclear species. 
F and H denote nonlinear functions of the arguments.
This system of equations is
solved implicitly from time $t^n$ to $t^{\rm n}=t^{\rm n}+\delta t$ according to
\begin{eqnarray}
Y^{n+1}_i - Y^n_i - \delta t \cdot F(Y^{n+1}_j,T^{n+1}) & = & 0 \\
\epsilon^{n+1} - \epsilon^n - \delta t \cdot H(Y^{n+1}_j,T^{n+1}) & = & 0
\end{eqnarray}
by a Newton-Raphson-solver.

The coupling to the hydrodynamic equations is done by solving this system
with $Y^n_j, T^n \mbox{ and } \rho^n$ after hydro time step n with step size
$\delta t$ is executed, and incorporating the 
results, namely $Y^{n+1}_i \mbox{ and } \epsilon^{n+1}$ into the hydrodynamic 
equations in the next time step n+1. Heat conduction, magnetic fields,
radiation and a possible rotation of the white dwarf are ignored.

Here, we use the same initial model for our two-dimen\-sional 
calculations as Glasner, Livne, \& Truran (1997).
Since for the problem of simulating subsonic convection in an explicit
scheme such as ours the time step (here: $\sim$ milliseconds)  
is limited by the cell size and 
the local sound speed, it is not possible to calculate 
by means of such schemes the whole
accretion process (several $10^5$ years) until the TNR takes
place. Moreover, even for an implicit code, such as VULCAN, solving the
full problem in 2-d is nearly impossible. Therefore   
the accretion phase and the slow stages of the burning were calculated by 
\cite{Glasner97} (1997) using a one dimensional implicit hydro code
(\cite{Glasner96} 1996). 

During this early phase hydrogen-rich matter of solar composition 
(Z = 0.02) is accreted onto
the surface of a 1 M$_{\odot}$ C+O 
white dwarf at a accretion rate of $5.0\cdot 10^{-9}$M$_{\odot}$yr$^{-1}$. 
As at the bottom of the envelope the temperature rises
convection sets in. Convection is taken into account by the
MLT in the one dimensional model. The model we obtained from A. Glasner and 
E. Livne provided the 
density, temperature, entropy, and composition profiles at the point at 
which the temperature at the bottom of the 
envelope had already reached 10$^8$ K. The mass of the hydrogen shell is
about $2 \cdot 10^{-5}$ M$_{\odot}$.
We mapped this model onto a radial row of our 2-d grid
and relaxed it for several hundred dynamical time scales and then mapped it
onto the full 2-d grid. The calculations were done with two
different resolutions of a domain  nearly identical to what has been used by 
\cite{Glasner97} (1997). This means that the grid covers 100 km of the outer part of the 
white dwarf and 1000 km of the envelope, radially, and $\sim$ 1800 km in the 
lateral direction, corresponding to the outer part of a 0.1$\pi$
segment in spherical coordinates. But
both grids are Cartesian and the curvature of the WD surface is
ignored for reasons given earlier. 
The coarser computational grid consists of 100 
uneven radial zones and 220 lateral zones which are equally spaced. Hence
the spatial resolution within the white dwarf and the first layers of the envelope
is about 5 km $\times$ 8 km in this case. The finer computational grid
consists of 500 uneven radial zones and 1000 lateral zones which are equally 
spaced. Here the spatial resolution within the white dwarf and the
first few layers of the envelope is about 1 km $\times$ 2 km. 

Our equation of state consists of the ideal Boltzmann gases of the
nuclei under consideration, an electron gas component with arbitrary
degeneracy and a photon gas component.

Finally, the boundary conditions and the initial perturbation 
have to be specified. Here we use a reflecting boundary at the bottom 
and an outflow boundary at the top of the grid. Periodic boundary conditions 
are applied laterally.
The model is initially perturbed by increasing temperature in one zone of
the bottom layer of the envelope by 1\% in both cases.
The further evolution is then followed for about 1400 physical seconds
in the case of the low resolution grid and for about 180 physical
seconds for the high resolution run.
In the next Section we summarize the results of both calculations and compare them. 

\section{Results}

\subsection{The low resolution simulation}

After the ignition of one zone the burning remains local for a 
few seconds (Fig.1b). Once the temperature has gone up 
a bit a first local TNR sets in. The burning front propagates
laterally with a velocity of about 15 km/s which is very close to the lateral
spread by small scale turbulence as was estimated by \cite{Fryxell82} (1982). This means that
it should take over 100 s to ignite the entire bottom layer of 
our computational domain by small scale turbulence. However, this argument
ignores the capability of sound waves to ignite an isothermal layer
which is close to the ignition temperature. In fact, we find that
spreading happens essentially on a sound travel time once a sufficiently
large volume burns and sound waves are emitted from this region,
and a full horizontal layer of the computational grid burns after 
about 14 s (Fig. 1c). 
Of course, our initial conditions, namely an isothermal layer very close to ignition
temperature, enhance this effect and, in reality, small scale turbulence may turn
out to be more important.
Already 20 s after ignition
the information about the location of 
the ignition point is largely lost. During the
following 500 s the mean energy generation rate in the
most violently burning shells is slowly increasing from 
about $10^{13}$ erg/g/s to over $10^{\rm 15}$ erg/g/s
(Fig. 2), and individual zones reach up to
$\sim 3 \cdot 10^{15}$ erg/g/s (Fig. 10b).

\begin{figure*}[t]
%\begin{tabular}{cc}
%\epsfxsize = 8.8cm
%\fbox{
%\epsfbox{Vel_low/NOVAcabe.ps}}
%&
%\epsfxsize = 8.8cm
%\fbox{
%\epsfbox{Vel_low/NOVAcaep.ps}} \\
%a) & b) \\ \\
%\epsfxsize = 8.8cm
%\fbox{
%\epsfbox{Vel_low/NOVAcafj.ps}}
%&
%\epsfxsize = 8.8cm
%\fbox{
%\epsfbox{Vel_low/NOVAcbam.ps}} \\
%c) & d) \\ \\
%\epsfxsize = 8.8cm
%\fbox{
%\epsfbox{Vel_low/NOVAccet.ps}}
%&
%\epsfxsize = 8.8cm
%\fbox{
%\epsfbox{Vel_low/NOVAchhg.ps}} \\
%e) & f) \\ \\
%\epsfxsize = 8.8cm
%\fbox{
%\epsfbox{Vel_low/NOVAcrma.ps}}
%&
%\epsfxsize = 8.8cm
%\fbox{
%\epsfbox{Vel_low/NOVAcujo.ps}} \\
%g) & h)
%\end{tabular}
\caption{\label{fig1} Velocity field at different 
stages of the evolution for the low 
resolution run. The color coding is done according 
to the absolute value of the velocity
at each point. T8 denotes the temperature of the hottest individual zone.}
\end{figure*}

In detail, the thermal history of the convectively burning shells 
turns out to be rather complicated. 
During the first 25 seconds cooling by convective mixing of cold envelope
material into the burning region is slightly  
more efficient than nuclear energy release.
Thus the maximum temperature drops a bit before rising
rather moderately,  and we do not find a change into a fierce 
runaway (Fig. 3a). This first result does not come as a
surprise. In fact, it is consistent with 1-d models which do not
assume initial enrichment of carbon and oxygen (see, e.g., 
Starrfield, Truran, Sparks, \& Kutter (1972), and Starrfield, Sparks,
\& Truran (1974))
and just reflects the fact that we do not find much mixing by
convective overshooting in the early phase of the TNR. The moderate increase 
of both maximum temperature and energy generation rate continues
until most of the protons are used up after approximately 500 s.

\begin{figure}[ht]
\epsfxsize = 8.8cm
\epsfbox{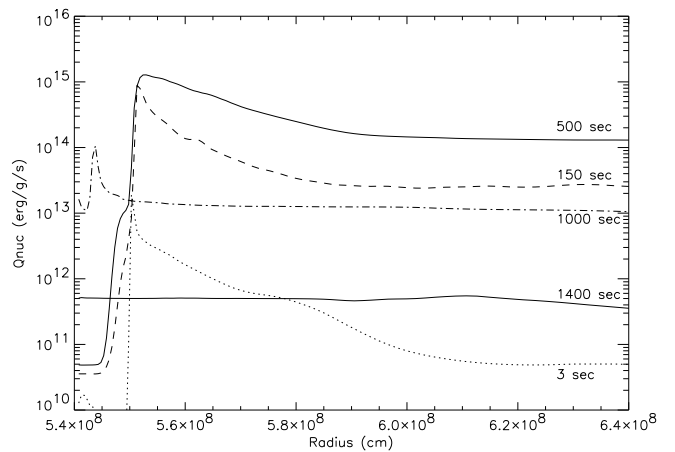}
\caption{\label{fig2} Laterally averaged vertical 
profile of the energy production
rate at several times of the low resolution calculation.}
\end{figure}

\begin{figure*}[ht]
\begin{tabular}{cc}
\epsfxsize = 8.8cm
\epsfbox{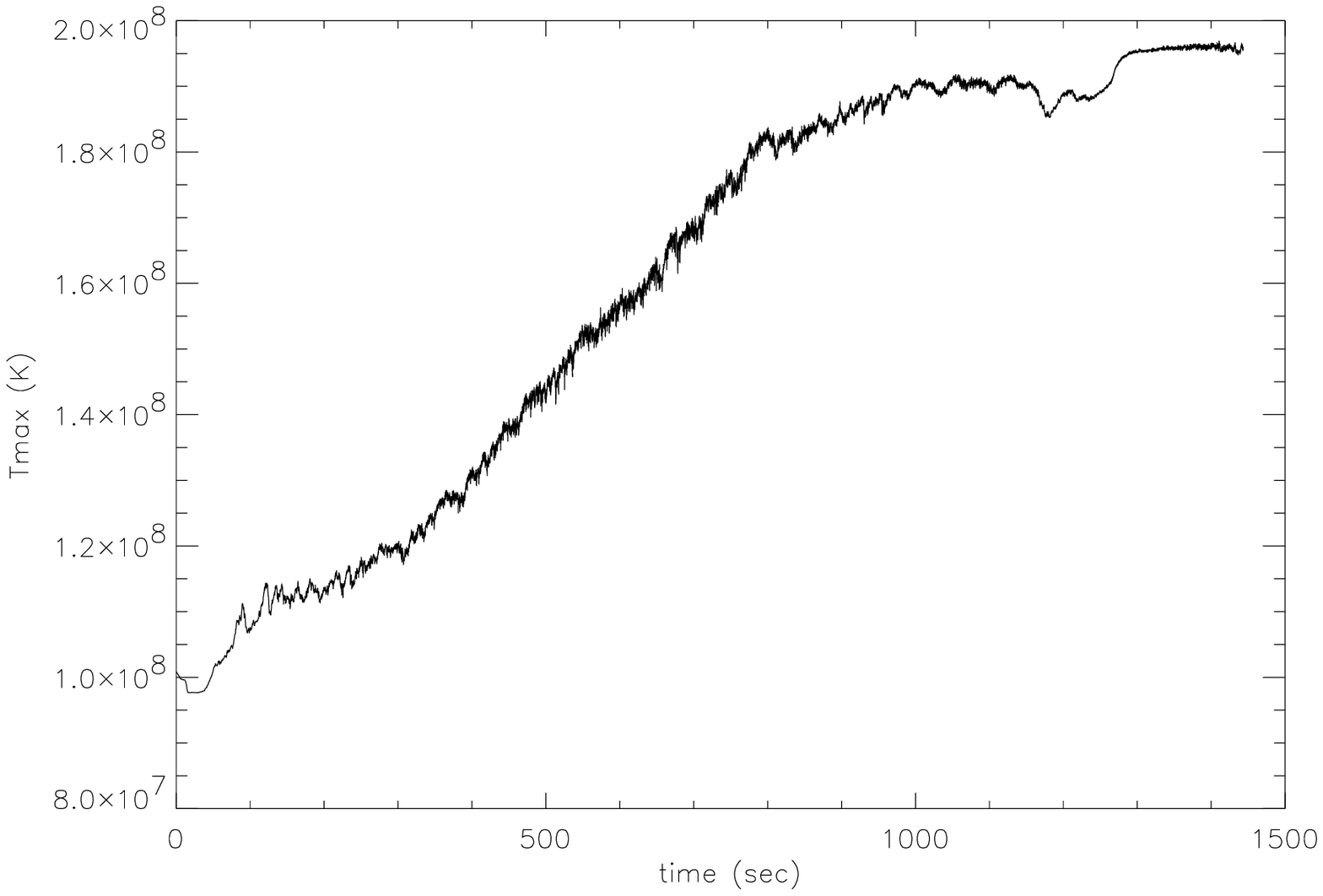}
 &
\epsfxsize = 8.8cm
\epsfbox{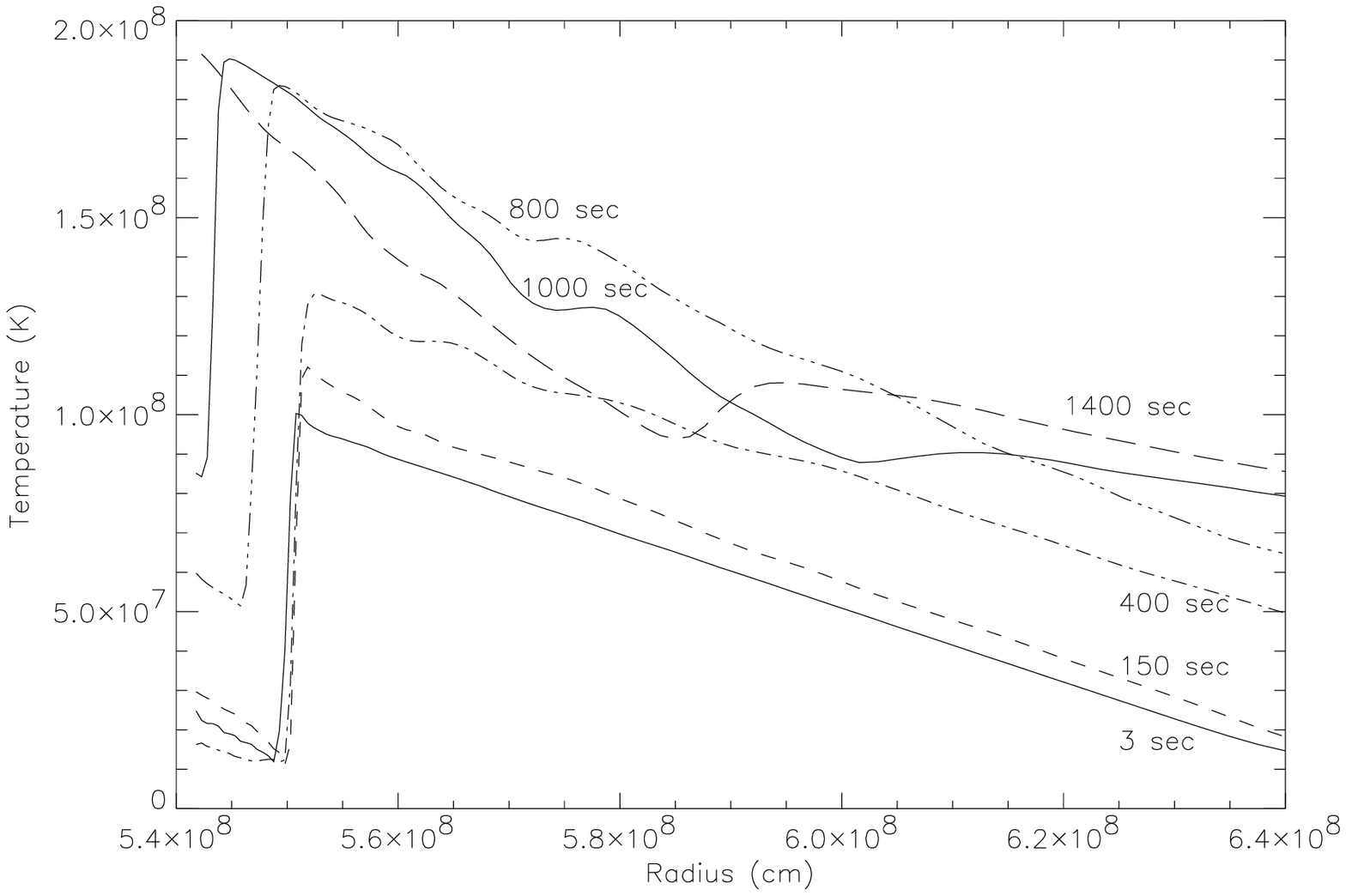} \\
a) & b)
\end{tabular}
\caption{\label{fig3} Temperature evolution of the
low resolution run. a) is the temperature
of the hottest envelope shell and b) the laterally
averaged vertical temperature profile
at several times.}
\end{figure*}

This behavior of our simulation, which is distinctly different from
earlier 2-d work (i.e., \cite{Glasner97} (1997)) can be understood as follows.
We find that during the first 100 to 150 seconds after ignition 
the flow field reaches a quasi-stationary state in which
the whole envelope is convectively stirred up and 
the dominant eddy size is about 200 km only (Fig. 1e). 
These eddies appear to be very stable. They rarely 
merge to larger convective structures but rather behave like particles 
which scatter off after collisions, in contrast to what one expects
to happen for homogeneous and isotropic 2-d turbulence. However, this
finding is not new but has already been discussed in the turbulence
literature much earlier (\cite{Fornberg77} (1977), \cite{Basdevant81} (1981), 
and McWilliams (1984)). 
There it was suggested that the long time evolution of turbulent 2-d 
flow fields is dominated by small coherent vortices. 

In order to clarify this point we will,
in the following, give a short summary of the theoretical
background of 2-d turbulence. In general, 2-d turbulence is the
study of solutions of the incompressible Navier-Stokes equation
which depend only on two cartesian coordinates, say x and y.
Since we solve the Euler equation, we confine the discussion
to the case when the viscosity $\nu$ is zero. According to 
\cite{Frisch95} (1995) the z component of the Euler equation
is a simple advection diffusion equation without back reaction
on the x-y plane. The vorticity has just a vertical component 
with respect to the x-y plane. Taking the curl of the Euler
equation the vorticity equation follows which has, under
the foregoing assumptions, just a z component which we denote
by $\zeta$. Introducing a stream function $\psi$ by $V_x=\partial_y \psi$
and $V_y=-\partial_x \psi$ the vorticity equation of the incompressible
Euler equation reads :
\begin{eqnarray}
\partial_t \zeta &=& J(\psi,\zeta) + \eta \nonumber \\
\Delta \psi &=& - \zeta,
\end{eqnarray}
where $J(\psi,\zeta)=\partial_x \psi \partial_y \zeta - \partial_x \zeta \partial_y \psi$
is the Jacobian Operator and $\eta$ is the curl of the external driving force.
This equation is vorticity conserving when viscosity (here of course $\nu = 0$)
and external forces are ignored. 

Because of the symmetry chosen we have $\vec{\nabla} \times \vec{\nabla} \phi = 0$
therfore $\eta=0$. However,
in our case we consider general compressible flows with energy sources and
also numerical viscosity comes into
play. 
As a consequence vorticity conservation is not exactly fullfilled. On the other hand, in
situations where incompressibility is a good approximation and the energy
source term is small vorticity is nearly conserved, and
in fact we do find such situations in our simulations.

If one assumes that vorticity is exactly conserved the arguments in favour 
of stable coherent vortices proceed as follows: The solutions of
\begin{equation}
J(\psi,\Delta \psi) = 0
\end{equation}
are a subclass of solutions to eq. (6) for $\eta=0$ in a suitable frame of 
reference where
the vorticity $\zeta$ is not explicitly time dependent.
This also includes axially symmetric circular vortices where the stream
function $\psi$ is just a function of the distance to the vortex center.

According
to \cite{Frisch95} (1995) 
cascade arguments
are not valid anymore because vorticity conservation guarantees
the stability of those vortices, provided they are approximately 
axially symmetric. As was pointed out
by \cite{Frisch95} (1995) such axi-symmetric structures behave 
like point vortices, as long as they are
well separated from each other, or like
drops of 'laminar' fluid in an otherwise turbulent flow.

In our simulations we indeed do find these quasi-stable vortices, as
long as the nuclear energy generation rate is moderate and the background
flow on large scales is subsonic, such that incompressibility is a good
approximation. They are
responsible for the fact that we do not observe much overshooting and
mixing during the first 150 s of the TNR. 
The maximum velocities of these eddies can get very close to 
sound speed ($\sim$ 1400 km/sec at 150 seconds) and may even exceed it
later on. So a large part of the convective (and turbulent) kinetic
energy is locked in structures which do not contribute
significantly to the mixing. In contrast, typical velocities
of the large scale background flow are a few 100 km/s at that time.
The characteristic scale of a few 100 km is explained by numerical
resolution effects  at the lower end and the distortions of the
axi-symmetry once their linear dimension approaches a pressure
scale-height. So we expect that the typical scale will get smaller
with increasing numerical resolution, and we indeed find this effect
as will be shown in the next subsection.

\begin{figure}[ht]
\epsfxsize = 8.8cm
\epsfbox{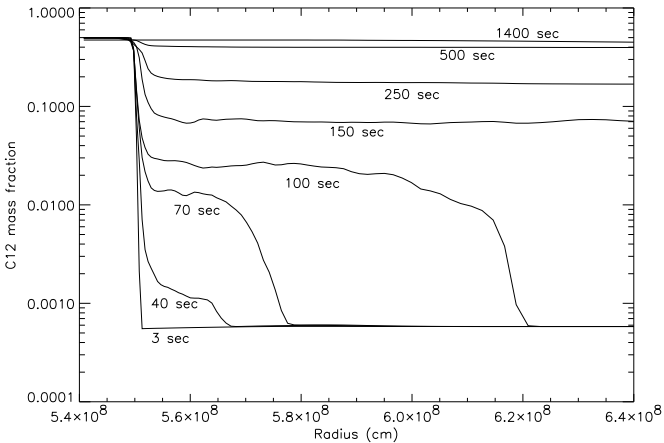}
\caption{\label{fig4} Laterally averaged vertical 
profile of the $^{\rm 12}C$ mass
fraction at several times of the low resolution calculation.}
\end{figure}

\begin{figure}[ht]
\epsfxsize = 8.8cm
\epsfbox{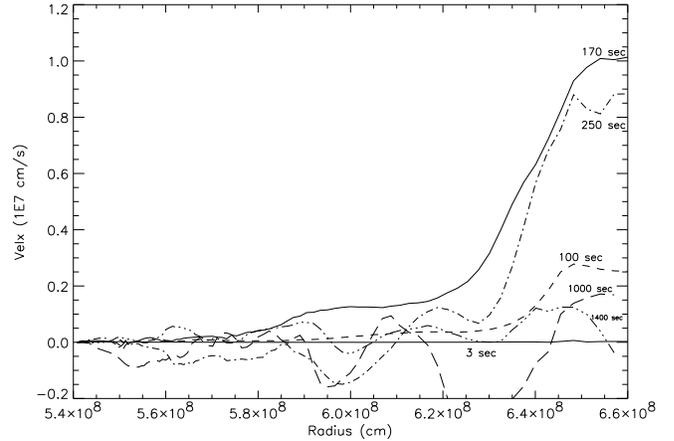}
\caption{\label{fig5} Laterally averaged vertical 
profile of the vertical component of the
velocity at several times of the low resolution calculation.}
\end{figure}

After about 150 seconds the most violently burning
shells reach nearly the peak nuclear energy generation rate
and stay approximately constant there (with very slow further 
increase) until the maximum is reached after about 500 seconds. 
During that phase the nuclear energy release still keeps rising gradually
in the outer parts of the envelope (Fig. 2) due to radioactive  
decay heating caused by the enhancement of $\beta^{\rm +}$-unstable
nuclei such as $^{\rm 15}$O there, but the small-scale vortices
gradually disappear. Our interpretation is that they are purely
a consequence of inertial range turbulence (which explains why their 
typical scale is always smaller than the pressure scale-height)
and they disappear once they are heavily distorted by nuclear energy
generation. Consequently, we start to see considerable mixing of white
dwarf core material into the envelope. While in the earlier phase 
(until 150 s) the mixing of C and O into the envelope resulted in 
an overall $^{\rm 12}$C mass fraction of only about 7$\%$ (Fig. 4), 
it now rises gradually to almost 40$\%$ at 500 s in that
part of the envelope not lifted off the white dwarf.  
  
The $^{\rm 12}$C mass fraction and the envelope metallicity are measures for the 
dredge up of core material,
since the CNO cycle does not lead to a higher number of 
CNO nuclei in the envelope but just distributes the shell metallicity
among the different CNO nuclei. The spatial distributions of $^{\rm 12}$C 
are shown in Figs. 10e and 10f at an early time (70 s) and 
later near the end of this phase (500s).  
The distribution looks rather smooth (note that in going from panel 
e to f the color coding was changed), showing that the convective times  
are indeed short as compared to the nuclear burning time scale. 

Also, envelope
matter is expelled with velocities of up to 100 km/s at 170 s
and beyond  (Fig. 5), which leads to a remarkable decrease 
of the envelope's density (Fig. 6). It decreases until 
minimum values are reached after 300 s (Fig. 6). From then on
mixing of core material into the envelope 
becomes very efficient, and C and O are mixed at a rate which 
leads to increasing the envelope density again. The flow field evolves 
from a state dominated by small eddies of scales
of 200 km to one which looks very inhomogeneous and incoherent,
dominated by large scale motions. Fig. 1f is 
representative for this phase, in which nuclear burning affects the
flow patterns considerably. 
  
During all of the phases discussed so far $\beta^{\rm +}$-unstable 
nuclei are produced in the burning region and are 
convectively spread over the entire envelope (Fig. 7, 10g,h). 
This is reflected in Fig. 2 by the nearly constant energy generation
rates as a function of distance from the core.
The horizontal tails seen in this figure are due to
energy release by temperature independent $\beta$-decays, 
especially $^{\rm 15}$O (Fig. 8) and $^{14}$O.
As long as the time scale for the production of $^{\rm 15}$O is still shorter
than its decay time we find an increasing mass fraction of
$^{\rm 15}$O, peak values being reached after roughly
800 s (Fig. 7). The fact that the temperature keeps
rising and the energy generation rate drops more slowly than would be
predicted from $^{15}$O decay alone is due to the formation and decay
of  $^{14}$O  which sets in once the envelope has been enriched
with core material (see Table 1). 

\begin{table}
\begin{center}
\begin{tabular}{|c|c|c|c|} \hline
Time & $^{14}O$ & $^{15}O$ & $^{17}F$ \\ \hline
800 sec & $1.0 \cdot 10^{14}$ & $5.3 \cdot 10^{11}$ & $2.8 \cdot 10^{11}$ \\ \hline
1000 sec & $1.7 \cdot 10^{13}$ & $1.6 \cdot 10^{11}$ & $4.7 \cdot 10^{10}$ \\ \hline
1200 sec & $2.7 \cdot 10^{12}$ & $5.0 \cdot 10^{10}$ & $6.5 \cdot 10^{9}$ \\ \hline
\end{tabular} \\ [0.5ex]
\end{center}
Table 1 : Energy production rate contributions in erg/g/s of different $\beta$-unstable
isotopes at different times calculated for the envelope shell at 6000 km. The dominant
energy production in the outer parts of the envelope is due to the $\beta$-decay of
$^{14}O$. 
\end{table} 

At this time  only a few protons are left in the envelope 
(Fig. 8) and the violent burning is close to being
extinguishing. The dredge-up of core material
yields a gradual removing of the outer parts of the white dwarf 
(Fig. 6) which becomes remarkable after about 1000 s.
The energy production by radioactive decays gets low and the
temperature increase levels off. Small scale vortices reappear, as
is to be expected from the previous discussion.

\begin{figure}[ht]
\epsfxsize = 8.8cm
\epsfbox{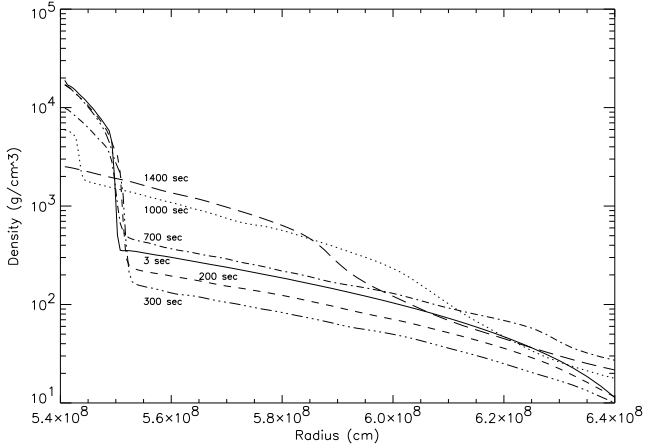}
\caption{\label{fig6} Laterally averaged vertical profile of the density
at several times of the low resolution calculation.}
\end{figure}

\begin{figure}[ht]
\epsfxsize = 8.8cm
\epsfbox{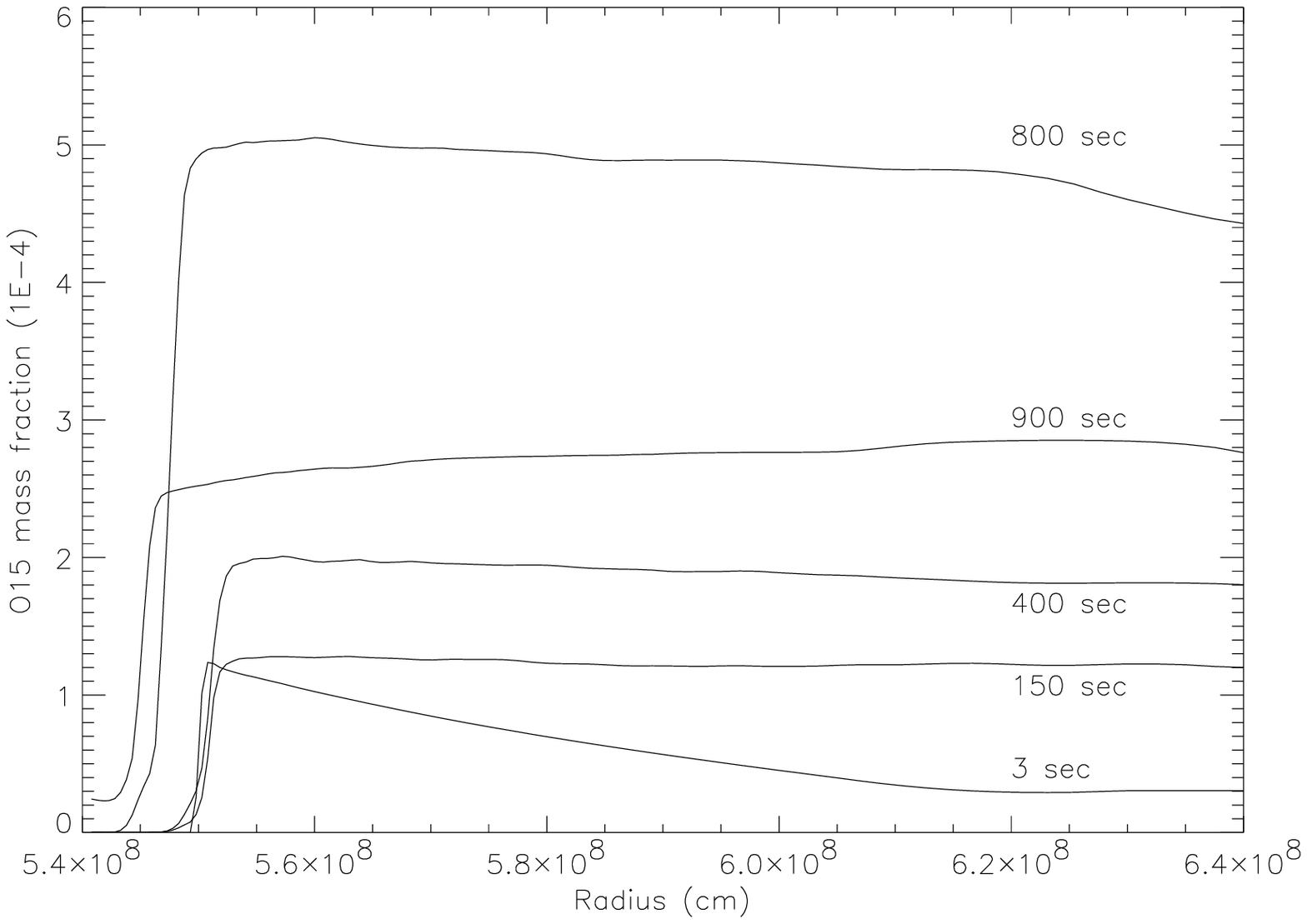}
\caption{\label{fig7} Laterally averaged vertical 
profile of the $^{\rm 15}O$ mass
fraction at several times of the low resolution calculation.}
\end{figure}

\begin{figure}[ht]
\epsfxsize = 8.8cm
\epsfbox{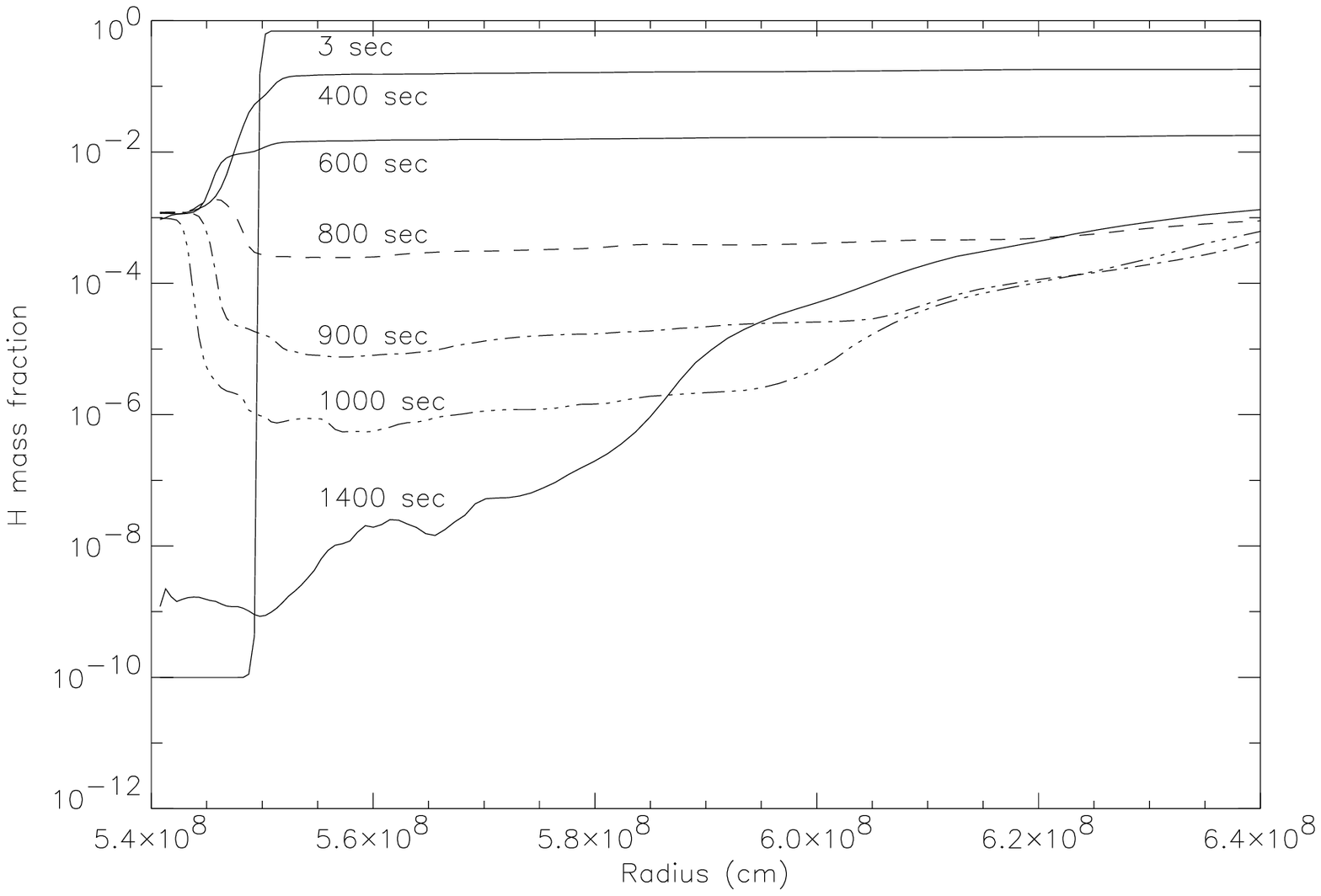}
\caption{\label{fig8} Laterally averaged vertical 
profile of the proton mass
fraction at several times of the low resolution calculation.}
\end{figure}

\begin{figure}[ht]
\epsfxsize = 8.8cm
\epsfbox{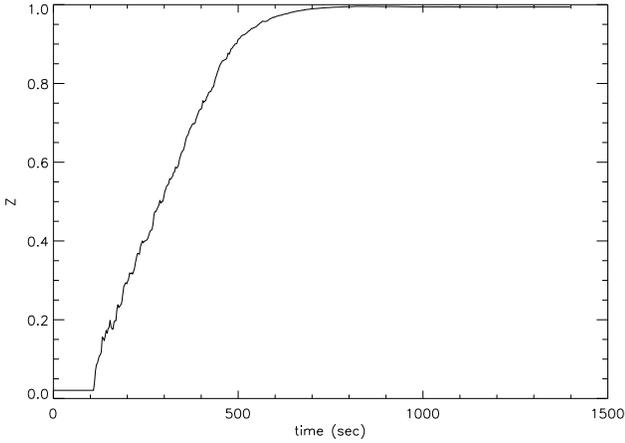}
\caption{\label{fig9} Evolution of metallicity 
for the last five shells of the
simulated envelope domain for the low resolution run.}
\end{figure}

Protons which have been mixed into the cool core during previous stages
(Fig. 8) and stored there and, due to low core temperatures,
have not been fused by the CNO-cycle are now mixed into the envelope
again. Here they are instantaneously absorbed via proton capture reactions  
almost exclusively in the hottest envelope shells resulting in a peak
of the energy production there (Fig. 2). 
Although the effect of this extra burning on the over-all dynamics is
not very big it may have interesting consequences for the
nucleosynthesis yields expected from the outburst, because a small
fraction of it might be mixed convectively all the way to the surface.

The growing envelope mass and the slowly decreasing nuclear energy 
release leads to a decrease of outflow velocities, 
until the envelope almost stops expanding near the end of the
computations (Fig. 5). Most of the outer core material
covered by the computational grid has been mixed into the 
envelope by that time (Fig. 6). The convective 
velocities decrease steadily (Fig. 1h) and the envelope
approaches a new hydrostatic equilibrium state.
Most of the previously stirred up heavy element enriched matter is 
settling onto the surface of the white dwarf at the end of 
the simulations (Fig. 6). The formation of a new core-envelope boundary 
is evident from the smooth density transition seen in Fig. 6 
at 1000 and 1400 seconds, respectively, which is also visible in other observables like 
temperature (Fig. 3b) or hydrogen mass fraction
(Fig. 8). 

Finally, in Fig. 9 we show how the metallicity  of the ejected
envelope material evolves with time.  In order to do so we estimate the
metallicity in the upper five calculated shells of the envelope and
assume that they will be ejected. The values we obtain in this way are
Z $\simeq$ 0.2 \% for the maximum outflow velocities around
170 s,  increasing up to about Z = 0.9 \% after 500 s 
when the outflow velocities are still remarkable.  

\begin{figure*}[t]
%\begin{tabular}{cc}
%\epsfxsize = 8.8cm
%\fbox{
%\epsfbox{All_low/NOVAcbam_q.ps}}
%&
%\epsfxsize = 8.8cm
%\fbox{
%\epsfbox{All_low/NOVAchhg_q.ps}} \\
%a) & b) \\ \\
%\epsfxsize = 8.8cm
%\fbox{
%\epsfbox{All_low/NOVAcbam_tem.ps}}
%&
%\epsfxsize = 8.8cm
%\fbox{
%\epsfbox{All_low/NOVAchhg_tem.ps}} \\
%c) & d) \\ \\
%\epsfxsize = 8.8cm
%\fbox{
%\epsfbox{All_low/NOVAcbam_c12.ps}}
%&
%\epsfxsize = 8.8cm
%\fbox{
%\epsfbox{All_low/NOVAchhg_c12.ps}} \\
%e) & f) \\ \\
%\epsfxsize = 8.8cm
%\fbox{
%\epsfbox{All_low/NOVAcbam_o15.ps}}
%&
%\epsfxsize = 8.8cm
%\fbox{
%\epsfbox{All_low/NOVAchhg_o15.ps}} \\
%g) & h)
%\end{tabular}
\caption{\label{fig10} Evolution of some of the 
quantities shortly after ignition (70 seconds)
and at peak energy generation rate (500 seconds).}
\end{figure*}

\subsection{The high resolution simulation}

Increasing the resolution of the grid by a factor of five in each spatial
direction means that computational costs increase by roughly a factor
of 125 (including the effect of a five times smaller CFL-time step). 
So it is obvious that we could not compute a model 
with this resolution to the end of
the TNR as was done for the previous run. However, since here we are
more interested in testing the numerical accuracy of our attempt to
simulate directly turbulent combustion in the accreted envelope
of a white dwarf rather then presenting very realistic models for 
a nova outburst which can be compared with observations 
(what in fact could be done and will be done in a forthcoming paper
since all the relevant physics is included in the code), we decided to go for  
the high resolution and to follow the highly resolved model as long as
necessary and possible. In practice this meant that we stopped the computations 
after a physical time of about 180 s, for reasons which will be
discussed later.
      
Figure 11 gives the velocity field at different times 
for the case of high resolution and should be compared with 
Fig. 1 of the low resolution run. 
We disturbed as before one zone with a temperature
amplitude of 1 \% . Because now the mass inside a single zone  
is much smaller, the energy perturbation is about 20 times smaller as
compared to the low resolution case. Naively one expects a somewhat
slower evolution in the beginning, and that is exactly what happens. We begin
with showing a first snapshot at 13 s (Fig. 9a) since then the 
distortion is comparable (in energy) to the low resolution case after
3 s.

\begin{figure*}[t]
%\begin{tabular}{cc}
%\epsfxsize = 8.8cm
%\fbox{
%\epsfbox{Vel_huge/NOVAcafb.ps}}
%&
%\epsfxsize = 8.8cm
%\fbox{
%\epsfbox{Vel_huge/NOVAcagf.ps}} \\
%a) & b) \\ \\
%\epsfxsize = 8.8cm
%\fbox{
%\epsfbox{Vel_huge/NOVAcalq.ps}}
%&
%\epsfxsize = 8.8cm
%\fbox{
%\epsfbox{Vel_huge/NOVAcapm.ps}} \\
%c) & d) \\ \\
%\epsfxsize = 8.8cm
%\fbox{
%\epsfbox{Vel_huge/NOVAcbbd.ps}}
%&
%\epsfxsize = 8.8cm
%\fbox{
%\epsfbox{Vel_huge/NOVAcbiy.ps}} \\
%e) & f) \\ \\
%\epsfxsize = 8.8cm
%\fbox{
%\epsfbox{Vel_huge/NOVAccgg.ps}}
%&
%\epsfxsize = 8.8cm
%\fbox{
%\epsfbox{Vel_huge/NOVAccnz.ps}} \\
%g) & h)
%\end{tabular}
\caption{\label{fig11} Velocity field at different 
stages of the evolution for the high
resolution run. The color coding is done according 
to the absolute value of the velocity
at each point. T8 denotes the temperature of the 
hottest individual zone.}
\end{figure*}

From then on the evolution of the TNR has many similarities but
also distinct differences to the previous one. 
It is evident from Fig. 11 that the high resolution run also 
yields small coherent vortices, but now on typical 
length scales of 40 to 50 km which are smaller compared to the low resolution
run by almost exactly the same factor by which the resolution was
increased, indicating that these structures are still not fully
resolved. Therefore, if it should turn out that the over-all
properties of the model are affected in a similar manner this would
mean that there is no chance to tackle the nova problem by means of
direct numerical simulations. Independent of this practical question, 
it is interesting to see that again 
the small eddies 
seem to carry a considerable amount of the turbulent kinetic energy. This 
finding gives us some hope that 2-d simulations can capture essential
features and that going to 3-d models might not be necessary, but, of
course, this conjecture has to be tested.   

The ignition phase is characterized by a flow field dominated by these
small scale vortices. Their typical velocities 
rise more rapidly. For instance the maximum velocities after 70 seconds for
the low resolution run are comparable to those of the high 
resolution run after 30 seconds. A likely interpretation is that now
the characteristic size of these eddies is considerably smaller than
the pressure scale height. Therefore they are less effected by
potential energy differences. Note that the fastest structures dominate
the color coding of Fig. 11 since we used a linear color scale.   

\begin{figure}[ht]
\epsfxsize = 8.8cm
\epsfbox{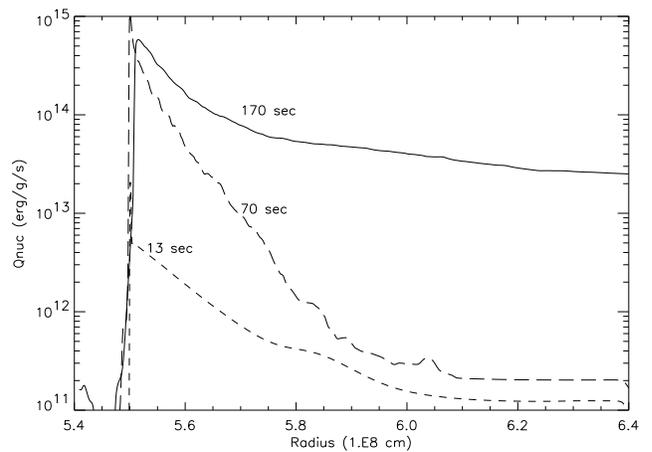}
\caption{\label{fig12} Laterally averaged vertical 
profile of the energy production
rate at several times of the high resolution calculation.}
\end{figure}

\begin{figure}[ht]
\epsfxsize = 8.8cm
\epsfbox{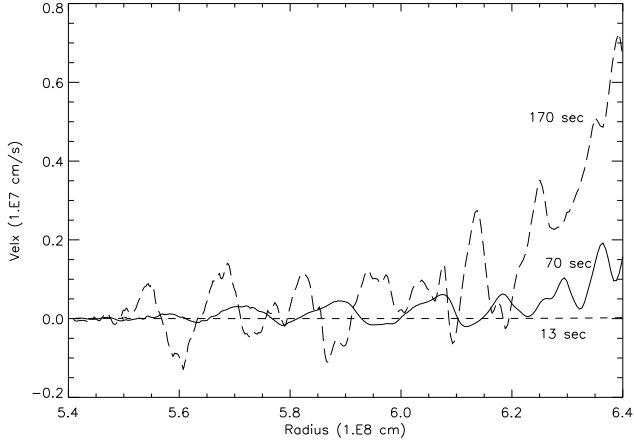}
\caption{\label{fig13}  Laterally averaged vertical 
profile of the vertical component of the velocity
at several times of the high resolution calculation.}
\end{figure}

According to Fig. 11a,b,c the lateral spread of the
burning by small scale turbulence along the white dwarf surface 
proceeds at a velocity of approximately 10 km/s during 
the ignition phase, very similar to what we have found in the  
low resolution case, and in accord with  
the estimates of \cite{Fryxell82} (1982). From then on 
again sound waves ignite the burning in  
other parts of the bottom shell of the envelope and they can also be
seen propagating vertically in Fig. 11d. After about 40 s the
entire bottom layer is burning. So the time required to propagate the
front is equal to what we have found  for lower resolution. This
result is in agreement with the sound wave interpretation because the
sound velocity is independent of the resolution. It also demonstrates
that at least during the ignition phase {\it lateral} energy transport 
is not dominated by the fast moving small structures. However, the 
{\it vertical} extend of the convective region is affected as can be seen by
comparing panel d of Fig. 1 with panel e of Fig. 11 but the
convective energy is mainly locked in small scale motions.

\begin{figure}[ht]
\epsfxsize = 8.8cm
\epsfbox{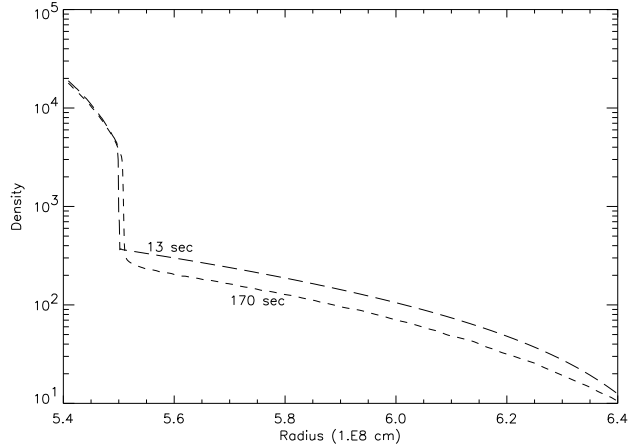}
\caption{\label{fig14}  Laterally averaged vertical 
profile of the density
at several times of the high resolution calculation.}
\end{figure}

\begin{figure}[ht]
\epsfxsize = 8.8cm
\epsfbox{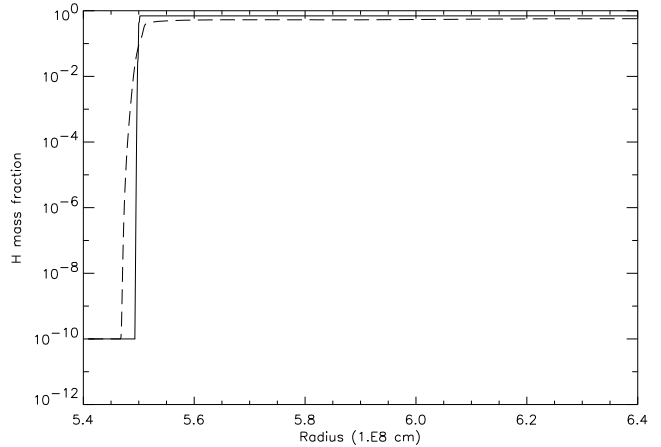}
\caption{\label{fig15} Laterally averaged vertical 
profile of the proton mass
fraction at several times of the high resolution 
calculation. Solid line : 13 seconds,
dashed line : 170 seconds.}
\end{figure}

\begin{figure}[ht]
\epsfxsize = 8.8cm
\epsfbox{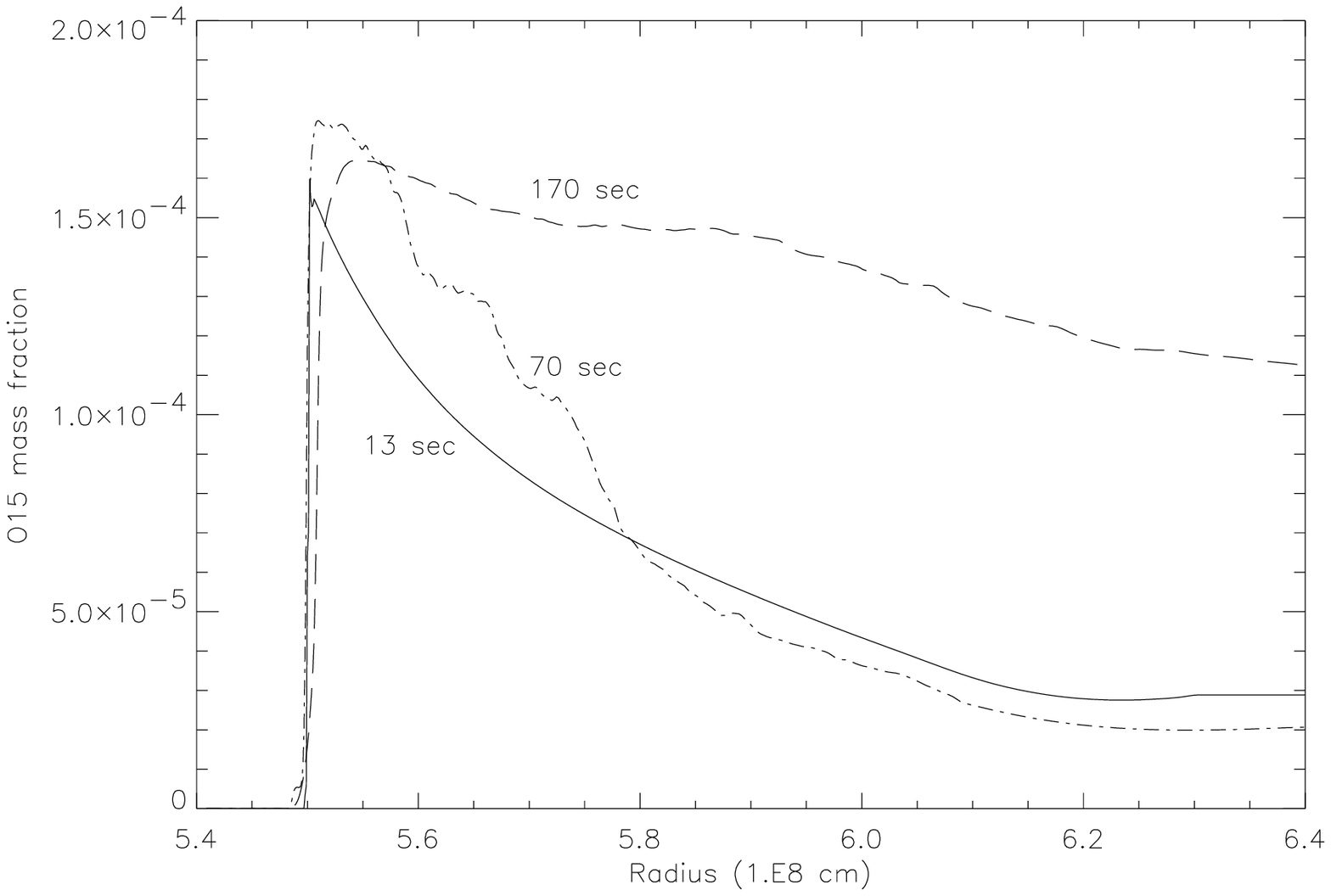}
\caption{\label{fig16} Laterally averaged vertical 
profile of the $^{\rm 15}O$ mass
fraction at several times of the high resolution calculation.}
\end{figure}

\begin{figure}[ht]
\epsfxsize = 8.8cm
\epsfbox{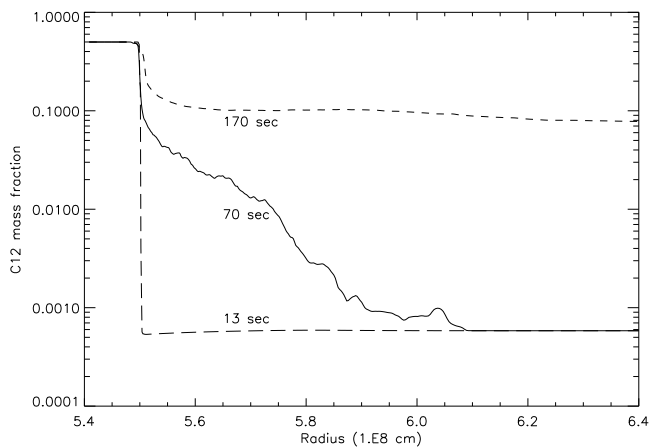}
\caption{\label{fig17}  Laterally averaged vertical 
profile of the $^{\rm 12}C$ mass
fraction at several times of the high resolution calculation.}
\end{figure}

As a result, we find that the temperatures in the hottest zones
increase a bit faster than previously (Fig. 12a), but the over all effect is
rather small as one can see by comparing the laterally averaged
temperatures given in Fig. 3b and Fig. 12b. 
This is a clear indication that larger scales are mainly responsible
for the energy transport, and they are well resolved in both
cases. The small delay of the temperature increase in the low
resolution run just reflects the fact that due to the lower velocities in the
small eddies more energy is available for the large scale motions and,
therefore, convective energy transport is a bit more efficient. It is
also obvious from comparing for example Fig. 2 and Fig. 12 or
Fig. 4 and Fig. \ref{fig17} that there are only minor differences between the
two calculations for key quantities such as the energy generation rate
or the $^{12}$C mass fractions once the envelope has become fully
convective after about 150 s. This, in fact, was the main reason
why we stopped the highly resolved run after about 180 s.

However, there is still a bit of uncertainty in this conclusion. It
appears that in the high resolution case the energy generation rate as
well as the maximum temperature reached in any of the shells seems 
to decrease or at least level off after about 150s. We belief that
this is a temporary effect only. As we have shown earlier, once the
energy generation rate has reached values around 10$^{15}$ erg/g/s
the small vortices begin to be destroyed by the more violent large
scale motions. It is very likely that the same will happen also in the
high resolution calculations, maybe with some delay, due to the higher
velocities (and therefore increased stability) in the smaller eddies.

Fig. 14 shows the laterally averaged
vertical distribution of the vertical velocity component. 
Not surprisingly, the high resolution run shows much 
more structure and rapid oscillation, especially still at 170 seconds,
whereas in the low resolution case the envelope is expanding more
uniformly at the time of peak outflow velocities (Fig. 5),
and they compare more with values reached already after 150 s in the
latter. But again, we consider this to be a minor uncertainty,
consistent with the general trend that for the better resolved
calculations the early phase of the TNR proceeds a bit faster, whereas
later, when larger eddies take over, the evolution is somewhat slower in
this case.
 
The density profiles (Fig. 15) show the decrease of the envelope
density. Just as before, consumption of protons also proceeds slowly
(Fig. 16). For completeness, we give the $^{15}$O
mass fraction and metallicity of the ejected shells in
Figs. 17 and 9, respectively. Of course, it would
have been nice to give all these data also at later times, but because
of the enormous computer time required for this run we stopped it.  

\section{Discussion and conclusions}

We have presented first results of numerical simulations of the TNR in
the hydrogen-rich envelope of an accreting white dwarf. These
simulations were performed by means of a modified version of a
standard PPM-type hydro code which is able to treat reactive
hydrodynamics in one, two or three spatial dimensions. The models
outlined in the previous sections were obtained in planar geometry,
the surface layers of the white dwarf and its envelope being mapped
onto a 2-d grid in Cartesian coordinates. The main intention of the
simulations was to investigate the question as to whether convective
overshooting can or will lead to self-enrichment of the envelope with
C and O from the underlying white dwarf's surface. 

The same question was recently addressed in a very similar
simulation by Glasner, Livne \& Truran (1997) who found in fact a positive result. 
According to their study, considerable dredge-up of white dwarf matter occurs
early in the TNR, already after about 100 s, leading to a violent
runaway between 150 and 200 s after the computations had been started
in 2-d, with a temperature of 10$^8$K at the bottom of the envelope.
 
Our results agree in certain respects with those of \cite{Glasner97} (1997),
and disagree in others. We agree with them on the general outcome,
namely that even with an initially solar composition of the accreted
matter most of the envelope is ejected during the outburst. We also
agree that the ejected material will be enriched with matter mixed in
from the white dwarf. On the other hand, the differences are
remarkable, given the facts that we started our calculations from
exactly the same initial conditions and that, in one case, even the
numerical resolution was nearly equal to theirs. 

In general terms, the outburst is much less violent in our
calculations as compared to theirs. This shows up in the longer
timescale for the TNR, the lower ejection velocities, the considerably
lower temperature near peak energy production, etc.

\begin{figure}[ht]
\epsfxsize = 8.8cm
\epsfbox{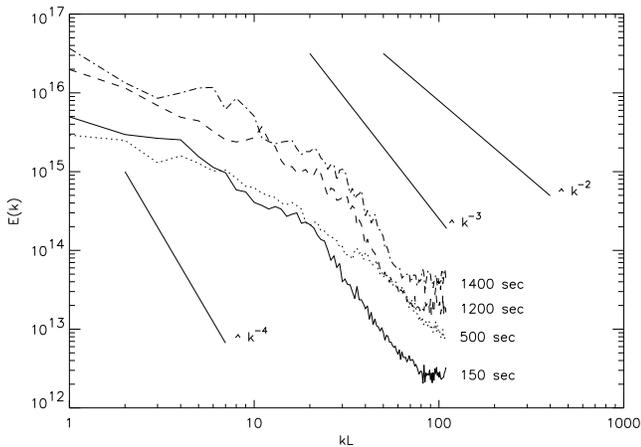}
\caption{\label{fig18} Energy spectra of the kinetic energy at different
times of the low resolution run. k is the wavenumber and L is the lateral 
gridlength. The energy is plotted in arbitrary units.}
\end{figure}

The cause of the deviations can be traced back to large differences
in the convective flow patterns we obtain.  In their simulations a few
large eddies dominate the flow already during early phases of the TNR,
whereas in our calculations small eddies carry a large amount of the convective
kinetic energy  until the burning proceeds at maximum rate.
In Fig. \ref{fig18} we provide energy spectra for the low resolution run,
to support this argument. They are not as steep as expected for instance for
unforced decaying incompressible turbulence as demonstrated in 
\cite{McWilliams84} (1984). 

Most of these eddies move very little in vertical direction and,
therefore, do not penetrate into the white dwarf. This is the reason
why we observe very little mixing until the envelope lift-off sets in
and dredge-up happens in the steep density (and pressure) gradient near
the white dwarf's surface. However, this late mixing has little effect
on the dynamical evolution of the envelope because at that stage most
of the hydrogen has already been burnt. 

If true, this finding
will, nevertheless, affect the nucleosynthesis predictions obtained 
from the model, but we are not 100$\%$ sure. The reason is that the
surface layers of the white dwarf are not well resolved in our ``low''
resolution run and, due to the Eulerian nature of the code numerical
diffusion may in part be responsible for some of the late mixing. In
principle this could be checked by extending the ``high'' resolution
run to about 800 s, but this would cost a lot of computer resources, and we
thought it might be more important to save those for 3-d simulations
which are under way.

The question then remains as to why the flow patterns are so different in
the two independent sets of calculations. Our interpretation is that
they are mainly caused by the different boundary conditions and
symmetries imposed. We think it is dangerous to impose polar symmetry
and reflecting boundary conditions in simulations which aim at
investigating geometrical properties of free subsonic convection. In fact,
we found in our simulations that the fast and small vortices moved
mainly horizontally all across the computational grid, leaving the
grid on one side and coming back from the other. Such motions
would be prohibited if reflecting boundaries were imposed, and it is
likely that such boundary conditions will suppress these
modes. Moreover, as can be seen in Fig. 1 and 2 of \cite{Glasner97} (1997)
in their computations most of the action appears to be either 
at the polar axis or near the horizontal
boundaries which indicates unphysical symmetries.

In conclusion, we have demonstrated that direct numerical simulations
of thermonuclear combustion in the accreted hydrogen-rich  envelopes
of white dwarfs are feasible. We have shown that even in 2-d (and in
contrast to common expectations) small scale structures carry a
large amount of
the convective kinetic energy. The smallest of those eddies 
were not resolved in our models but their actual size seems to 
be of minor importance for the general behavior of the TNR. Just the
fact that they exist makes a big difference, as the comparison with
the work of \cite{Glasner97} (1997)  showed. They were interpreted as stable
solutions of the (time independent) Euler equations which form from
small fluctuations in 2-d convective flows. The fact that they
disappeared once the energy generation (source terms) began to
dominate over the advection strongly supports this interpretation. 

It will be interesting to see which of these conclusions remain valid
in 3-d simulations. The small scale vortices might not be as important
any more because their stability rests largely on axi-symmetry which
is broken in 3-d. On the other hand side, their main effect was to
hinder the formation of dominant large scale eddies which otherwise
would show up more clearly in 2-d. So there is hope that 2-d simulations 
do capture the main features of the TNR, but this has to be seen.
\newline

{\it Acknowledgements.} The authors are grateful for many enlightening
discussions with Stanford E. Woosley, Jens C. Niemeyer and Ulrich Kolb. 
They thank Ami Glasner and Eli Livne for supplying the initial model,
Ewald M\"uller for an earlier version of the PROMETHEUS code, and Rudi
Fischer and Jakob Pichlmeier for their help in preparing a parallel version
of the code. They also thank the stuff members of the Rechenzentrum Garching
for their support during the computations. This work was supported in part 
by NSF at the Institute for Theoretical Physics at the University of 
California, Santa Barbara under Grant No. PHY94-07194 and by NASA under 
Grant NAG 5-3076 at the University of Chicago.
The computations were performed at the Rechenzentrum Garching on a
Cray T3E.


\begin{thebibliography}{99}

\bibitem[Basdevant, Legras, Sadourny \& Bel\'{a}nd]{Basdevant81} Basdevant, C., 
Legras, B., Sadourny, R., and Bel\'{a}nd, M.,
1981, J. Atmos. Sci., 38, 2305 

\bibitem[Colella \& Woodward]{Colella84} Colella, P., and Woodward, P.R.,
1984, J. Comp. Physics, 54, 174

\bibitem[Fornberg]{Fornberg77} Fornberg, B., 
1977, J. Comp. Physics., 25, 1

\bibitem[Frisch]{Frisch95} Frisch, U., 
{\em Turbulence}, Cambridge University Press 1995 

\bibitem[Fryxell \& Woosley]{Fryxell82} Fryxell, B.A., and Woosley, S.A.,
1982, ApJ, 261, 332

\bibitem[Fryxell, M\"uller \& Arnett]
{Fryxell89}  Fryxell, B.A., M\"uller, E., and Arnett, W.D.,
1989, Max-Planck-Institut f\"ur Astrophysik Report 449, Garching, Germany 

\bibitem[Glasner \& Livne]{Glasner95} Glasner, S.A., and Livne, E., 1995,
ApJ, 445, L149

\bibitem[Glasner,  Livne \& Truran]{Glasner95_2} Glasner, S.A., Livne E.,
and Truran, J.W., 1995, in IAU Colloq. 158, Cataclysmic Variables and Related
Objects, ed. E. Evans \& J.H. Wood (Kluwer: Academic), in press

\bibitem[Glasner \& Truran]{Glasner96} Glasner, S.A., and Truran J.W., 1996,
in preparation

\bibitem[Glasner, Livne \& Truran]{Glasner97} Glasner, S.A., and Livne E.,
and Truran, J.W., 1997, ApJ, 475, 754

\bibitem[Livne]{Livne93} Livne, E., 1993, ApJ, 412, 634

\bibitem[Mac Donald]{MacD80} Mac Donald, J., 1980, MNRAS, 191, 933

\bibitem[McWilliams]{McWilliams84} McWilliams, J.C.,
1984, J. Fluid Mech., 146, 21

\bibitem[M\"uller]{Mueller86} M\"uller, E., 1986, A\&A, 162, 103

\bibitem[M\"uller]{Mueller94} M\"uller, E., 1994, in {\it
Supernovae} (Proc. Les Houches Summer School 1991),
ed. S.A. Bludman, R. Mochkovitch, and J. Zinn-Justin
(Amsterdam:Elsevier), p. 395

\bibitem[M\"uller]{Mueller97} M\"uller, E., 1997, in {\it
Gravitational Radiation} (Proc. Les Houches Summer School 1995), ed. 
S. Bonazzola and J.-A. Marck (Cambridge:Cambridge Univ. Press), in
press

\bibitem[Niemeyer \& Hillebrandt]{Niemeyer95} Niemeyer, J.C., and Hillebrandt, W., 1995,
ApJ, 452, 779

\bibitem[Prialnik, Shara \& Shaviv]{Prialnik78} Prialnik, D., Shara, M.M.,
and Shaviv, G., 1978, A\&A, 62, 339

\bibitem[Shankar, Arnett \& Fryxell]{Shankar92} Shankar, A., and Arnett, W.D.,
Fryxell, B. A., 1992, ApJ, 394, L13

\bibitem[Shankar \& Arnett]{Shankar94} Shankar, A., and Arnett, W.D., 1994,
ApJ, 433, 216

\bibitem[Shara]{Shara81_82} Shara, M.M.,
1981, ApJ, 243, 926, 1982, ApJ, 261, 649

\bibitem[Spiegel]{Spiegel63} Spiegel, E.A., 1963, ApJ, 138, 216

\bibitem[Starrfield]{Starr89} Starrfield, S., 1989, in {\em The Classical Novae},
ed. M. Bode and A. Evans, (Wiley: NY), 39

\bibitem[Starrfield]{Starr93} Starrfield, S., 1993, in {\em The Realm of Interacting
Binary Stars}, ed. J. Sahade, G.E. McCluskey, and Y. Kondo (Dordrecht: Kluwer), 209

\bibitem[Starrfield]{Starr95} Starrfield, S., 1995, in {\em Physical Processes in
Astrophysics}, ed. I. Roxburgh, and J.L. Masnou, (Springer: Heidelberg), 99

\bibitem[Starrfield, Sparks \& Truran]{Starr74} Starrfield, S., Sparks, W.M.,
and Truran, J.W., 1974, ApJ Suppl., 28, 247, 1985, ApJ, 291, 136

\bibitem[Starrfield, Truran, Sparks, \& Kutter]{Starr72} Starrfield,
S., Truran, J.W., Sparks, W.M., and Kutter, G.S., 1972, ApJ, 176, 169
 
\bibitem[Thielemann]{Thielemann96} Thielemann, F.-K.,
1996, private communication

\bibitem[Truran]{Truran82} Truran, J.W., 1982, in {\em Essays in Nuclear
Physics}, ed. C.A. Barnes, D.D. Clayton and D.N. Schramm (Cambridge: Cambridge
U. Press), 467

\bibitem[Truran]{Truran90} Truran, J.W., 1990, in {\em The Physics of Classical
Novae}, ed. A. Cassatella and R. Viotti, (Heidelberg: Springer), 373

\bibitem[Wallace \& Woosley]{Wallace81} Wallace, R.K., and Woosley, S.E.,
1981, ApJS, 45, 389

\bibitem[Woosley]{Woosley86} Woosley, S.E.,
1986, in {\em Nucleosynthesis and Chemical Evolution}, 
ed. B. Hauck, A. Maeder, and G. Magnet (Sauverny:Geneva Observatory)

\end{thebibliography}
\end{document}